\documentclass[11pt,twoside]{article} 
\usepackage{asp2004}
\usepackage{epsf}
\usepackage{psfig}
\usepackage{lscape} 

\begin{document} 
\title{Searching for White and Brown Dwarfs in the frame of the 
MUSYC/CYDER survey}
 \author{M. Altmann$^1$, R.A. M\'endez$^1$, M.-T. Ruiz$^1$, W. van Altena$^2$, E. Gawiser$^2$, J.Maza$^1$, and P. van Dokkum$^2$} 
 \affil{$^1$Departamento de Astronomia, Universidad de Chile, Casilla 36-D, Correo Central, Santiago, Chile\\
$^2$Department of Astronomy, Yale University, P.O. Box 208101, New Haven, CT 06520-8101, USA \\}
\begin{abstract} 
MUSYC ({\bf MU}ltiwavelength {\bf S}urvey by {\bf Y}ale and {\bf C}hile),
 and its predecessor, the CYDER ({\bf C}al\'an {\bf Y}ale {\bf D}eep
{\bf E}xtragalactic {\bf R}esearch) survey are both deep multiwavelength
surveys mainly aiming at extragalactic issues; however they were both designed
with a strong Galactic component in mind. Therefore they consist not only of
multipassband imaging ($UBVRIzJK$) and wide field multi object spectroscopy - 
it is also foreseen to create a multi-epoch dataset, 
thus allowing proper motions
to be derived using data of 1 and 4-5 years baseline. This enables us to
identify fast moving objects, and derive tangential velocities of nearby and 
intrinsically faint objects, such as White Dwarfs 
(especially cool WDs) and Brown
Dwarfs (BDs). For the former, we will be able to study the faint end
of the halo-field luminosity function and determine the fraction of baryonic
dark matter consisting of WDs. We will be able to put constraints on the volume
density of BDs, and determine whether this density actually meets expectations.
Here we give a report of work in progress. 
\end{abstract}

\section{Introduction}

With the advent of large CCD mosaic detectors it has become feasible to 
undertake surveys of selected parts of the sky of various field size, depth, 
and spectral coverage, some including spectra and/or proper motions. In 
empty, dust free fields of moderate to high Galactic latitude, which are mostly 
the subject of these surveys, most objects to interest are faraway galaxies
and galaxy clusters. However there are plenty of stellar objects which are 
of interest to Galactic and stellar astronomy. Among these are old objects,
such as White Dwarfs (WDs), Horizontal Branch Stars of various types or Pop II
Subdwarfs (metalpoor main sequence stars), and faint but nearby stars, e.g. M
Dwarfs and Brown Dwarfs (BDs). We will be able to detect BDs up to a distance of 
100 pc using proper motions and NIR photometry. White Dwarfs (see e.g. 
Pauli et al., 2003), especially the cooler objects, are important tracers of
the older populations of our Galaxy, i.e. Thick Disk and Halo. With this
study, we will be able to study the faint end of the Halo WD luminosity
function, by using the oldest WDs in the data set. This will allow us to 
constrain the fraction of WDs in baryonic dark matter, a long standing mystery.
Other hot stars - intrinsically much brighter than WDs - may help us to 
contrain the extent of the Galactic Halo. 
\begin{table}[!ht]
\caption{Celestial and Galactic coordinates and interstellar extinction of the
selected MUSYC (first four lines) and CYDER (lower part) fields}
\smallskip
\begin{center}
{\small
\begin{tabular}{lccrrc}
\tableline
\noalign{\smallskip}
 Field & RA &DEC  & $b$ & $l$ & $E_{B-V}$\\
          &hh:mm:ss& $^\circ:':''$ & $^\circ$ & $^\circ$ & mag\\
\noalign{\smallskip}
\tableline
\noalign{\smallskip}
CDF-S        & 03:32:29 & $-$27:48:47 & 224 & $-$54 & 0.01\\
SDSS 1030    & 10:30:27 &   +05:24:55 & 239 &   +50 & 0.02\\
Cast 1256+01 & 12:55:40 &   +01:07:00 & 306 &   +64 & 0.01\\
HDFS-Ext     & 22:32:35 & $-$60:47:12 & 328 & $-$49 & 0.03\\
\tableline
CYDER-A2     & 00:36:58 & $-$34:41:07 & 321 & $-$82 & 0.01\\
CYDER-A4     & 02:16:37 & $-$40:28:50 & 255 & $-$68 & 0.01\\
CYDER-D3     & 03:37:44 & $-$05:02:39 & 238 & $-$44 & 0.04\\
CYDER-C5     & 11:30:07 & $-$14:49:27 & 276 &   +44 & 0.04\\
CYDER-C2     & 12:52:48 & $-$09:24:30 & 305 &   +53 & 0.04\\
CYDER-C3     & 14:00:00 & $-$10:00:00 & 330 &   +49 & 0.04\\
CYDER-D1     & 22:01:38 & $-$31:46:30 &  16 & $-$54 & 0.02\\
CYDER-A1     & 22:44:22 & $-$40:07:49 & 359 & $-$61 & 0.01\\
\noalign{\smallskip}
\tableline
\end{tabular}
}
\end{center}
\end{table}

For the Galactic part of this project, we employ all four $30'\times30'$
MUSYC fields, as well as a selection of the preceding CYDER survey (see Table 1).
The former have full $UBVRIzJK$ coverage, but are more recent than the CYDER fields
(which mostly have $UBVI$); these allow us to determine 
proper motions accurate to $\sim$5 mas/yr. The MUSYC fields have
a baseline of (currently) only 1-2 years.
 This forces us to use two different approaches to
retrieve BD/WD candidates. For MUSYC, we preselect targets
using multi-colour photometry to be followed up by multi object
spectroscopy - proper motions will be added later to determine the
kinematics (``Spectroscopy first'', see Sect. 2.1). For CYDER, we first
isolate the high proper motion targets, and then classify them using
spectra and photometry (``Proper motions first'', see Sect. 2.2).

The optical imaging data was mostly obtained with the 4 m Blanco+MOSAIC II telescope
on Cerro Tololo, supplemented by data taken with WFI (ESO, La Silla), and
KPNO, the MOS  spectroscopy is being done at Magellan/IMACS and VLT/FORS+VIMOS,
the NIR imaging with Blanco+ISPI and the DuPont telescope on Las Campanas.
Source extraction, photometry and star/galaxy separation was done
using SExtractor (Bertin  \& Arnouts 1987). 
\section{Strategy}

\begin{figure}[ht!]
\plottwo{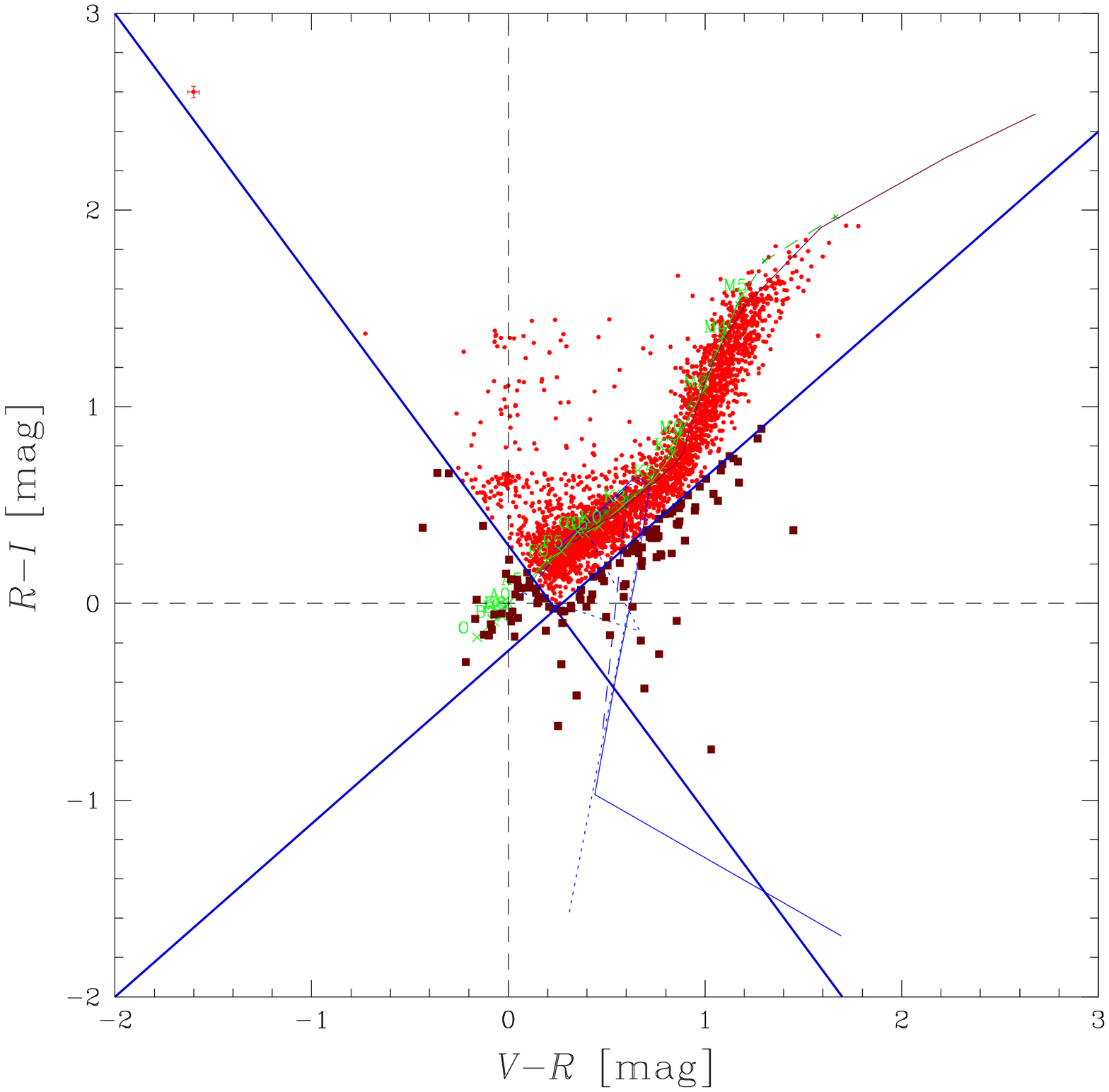}{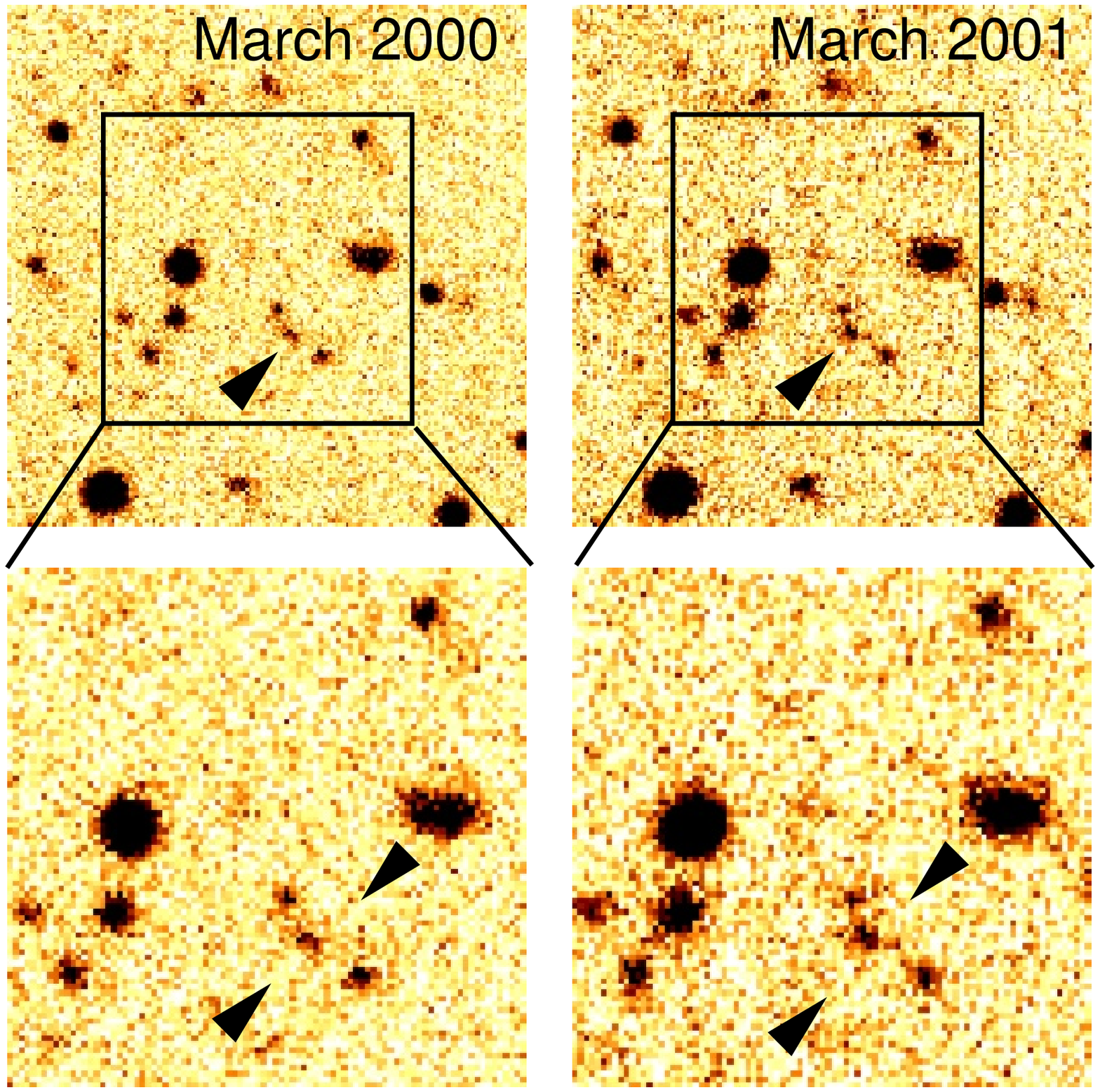}
\caption{Example of colour preselection (field HDFS-E). Note the prominent
distribution of main sequence stars. This plot shows two colour
selections. The line going from the upper left to the lower right 
separates hot stars, such as WDs and BHB/EHB stars; the other 
isolates cool WD candidates. The selected stars are depicted by bigger dots.
Also shown are WD cooling tracks and a template main sequence. }
\caption{One of the faint ($V\simeq26$ mag) fast moving objects found by image blinking
in field CYDER-C2. The images were taken $\sim$1 year apart; a crude
estimate leads to a proper motion of 0.5'' for this particular object.}
\end{figure}

\subsection{Spectroscopy first}
For some fields, especially the MUSYC fields, the observations
of multi object spectra has already started, long before we could think of 
deriving proper motions. Therefore our WD/BD candidates were preselected 
using multi-passband photometry. In order to keep the number of false candidates
as low as possible while keeping the completeness as high as possible, our
methods of preselection are being continuously refined. Employing the CLASS
parameter of SExtractor, we selected a sample of point sources. In
the beginning, we used simple colour cuts, such as $B-V<0$ to isolate blue
objects from the sample. However this still keeps a large number of 
misidentifications in the selected subsamples. A second step is to select
samples of targets using colour trends, as can be seen in Fig. 1. This
figure shows two different selections, one subsample of hot stellar targets, another one of
cool WD candidates.  Also shown are the main sequence (e.g. Baraffe et al. 1997,
and taken from Landolt Boernstein Vol. 2) and WD cooling tracks (Chabrier et al. 2000).
This already gives us a high confidence in most samples. Additional methods,
which are under consideration use fluxes in multiple passbands of template
model and standard star spectra to make the preselection, either by fitting
the templates to the measured spectra (Hatziminaoglou et al. 2000) or by using neural
 networks (Willemse, priv. comm.). Finally, the derivation of
proper motions, to gain access to the full kinematics if possible,
 will be done after the identification process.

\subsection{''Astrometry first'': Proper motions}
  For the proper motions we need data from at least two epochs. The datasets of MUSYC/CYDER are almost ideal for our purposes,
 because, they are 3 epoch data, with one baseline being short (i.e. 1 year), 
and the other significantly longer (4-5 years). For the CYDER fields the 
third epoch is currently being obtained. The short baseline allows us to 
account for high proper motion
 targets, i.e stars that move faster than ~0.25"/yr - when using longer 
baseline data these objects could be lost during the registration.
  
   However there might be  objects with proper motions in excess of 1" (which 
corresponds to 4 pixels in our data) - these might
 even be missed using the 1 yr baseline. To isolate these, we use several 
methods, one of which is the classical blinking of the two
 epochs by eye. The field CYDER-C2 has been blinked, revealing about 20 mostly very
faint objects with detectable motion, one of which is shown in Fig. 2.
The 3rd epoch will be needed to verify these
proper motions and to rule out other effects, which could mimic such motions,
such as distant supernovae, etc. Unfortunately, given their faintness, 
most of them can not be observed spectroscopically. Therefore 
photometry will have to provide us the informations needed to classify these 
elusive objects. The combination of faintness and large proper motion    
restricts the number of possible object types to two intrincially faint
types of objects: White and Brown Dwarfs. The latter will reveal itself by the 
very red colours, recently formed WDs due to their blue colour, and old WDs
due to their rather unique  spectral energy distribution.   

  The final accuracy of the proper motions (using all three epochs) is 
estimated to be in the order of 5 mas/yr. The nominal margins - as shown in 
an analysis of data from the 2.5 m DuPont telescope (which has an almost 
identical image scale)  are even smaller for all but the faintest stars - 
however we keep to this rather conservative estimate. 

\section{Discussion and outlook}
  Upto now we have the result of one spectroscopic campaign (CYDER-D1 \& D3). 
We found about 25 
mostly late M stars, and several hot stars, some even of spectral type B,
 as can be seen by the presence of He-lines. Since many spectra 
were of relatively low S/N it is hard to tell at first glance, some could 
be HB-like stars. Since the target selection of this 
campaign was done with the most primitive method we will be more efficient
in later campaigns.
  The field CYDER-C2 has been checked for objects of extremely high proper 
motion - about 20 candidates were found. 
%
This can only be a report on work in progress: However, with the spectroscopy 
of the MUSYC fields being well 
underway, and the 3rd. epoch observations of the CYDER fields coming up 
soon\footnote{Unfortunately the first campaign to secure the 3rd epoch data in 
August 2004 was lost completely due to inclement weather conditions.} 
we expect to be able to come to more significant results in the very near 
future.

\acknowledgements{MA, RM, MTR, acknowledge support from FONDAP 1503 0001
EG, from Fundacion Andes and from NSF Grant. No. AST-0201667. Special thanks to 
Roberto Antezana for his help with the plate blinking.}

\end{document}